# A comparative study of Different Machine Learning Regressors For Stock Market Prediction

Nazish Ashfaq[1,]
Punjab University Lahore

Muhammad Ilyas[1,2]
Univeristy Paris Est Cretiel
31 rue sedaine 75
11 Paris France
Email: Muhammad.ilyas@u-pec.fr

Dr. Zubair Nawaz[2]
Department of data science
Punjab University Lahore

*Abstract*—
**For the development of successful share trading strategies, forecasting the course of action of the stock market index i s important. Effective prediction of closing stock prices could guarantee investors attractive benefits. Machine learning algorithms have the ability to process and forecast almost reliable closing prices for historical stock patterns. In this article, we intensively studied the NASDAQ stock market and targeted to choose the portfolio of ten different companies belongs to different sectors. The objective is to compute the opening price of next day stock using historical data. To fulfill this task nine different Machine Learning regressor applied on this data and evaluated using MSE and R2 as performance metrics.**

## I. Introduction

For traders and analysts, stock market prediction is certainly one of the most difficult problems. People try to purchase shares in less price and sell them at high price, the difference in sale and purchase price is considered as surplus. Investors consider this business proficient and portable due to its liquid nature. Companies sell their shares to collect capital for new projects, while some companies allocate a fragment of their capital, to invest in other company's stocks. The whole purpose of this market is earning a profit. To increase profit, an investor tries to predict the future trends of the market that's why financial forecasting is one of the hottest topic these days. Almost all big organizations are hiring data scientist and financial analyst and they cannot predict the exact numbers, however, they can reliably predict the trends.

According to the fundamentals of the stock market, trends prediction depends highly on historical time series data. A time series is a series of numerical data points in consecutive order. It tracks the movements of the selected data points over regular intervals of time. The stock market generates this data on daily basis.

Time series analysis is the process of developing models with the help of different statistical techniques to represent underlying characteristics of time series data. This process generally involves in forging assumptions about the nature of data and then further decomposing it into constitution components. This work is important in a way as it can be used to evaluate the historical time series data in order to predict the future trends. The investors can make better decisions by using these historical trends.

In this study, nine different Machine learning (ML) algorithms are used for time series analysis of stocks data. Our proposed work uses Support Vector Regressor (SVR), Linear Regression (LR), Ridge Regression (RR), Lasso Regression (LR), ElasticNet (EN) Decision Tree Regressor (DTR), Random Forest Regressor (RFR), Extra Tree Regressor (ETR) and RANSAC.

Distribution of the remaining paper is as follows. Section two consists of previous work performed in time series data of stocks. Section three describes the dataset preparation while methodologies are discussed in section four. Evaluation of models is done in section five and discussion on results is a part of section six. Section seven concludes the paper along with limitations and future work.

## II. Back Ground

### A. Related Work

Prediction of financial market trends is a contradiction of the basic theory of finance, the Efficient Market Hypothesis(EMH). According to this hypothesis if one predicts the future trends using historical data he might not get any returns if the whole market notices this trend and as a consequence, the market corrects itself [9]. Many researchers have already rejected this hypothesis using various indicators and algorithms.

Stock price prediction is an important objective in the domain of finance [15] [16]. **Shetaet al.** in 2015 performed a comparison of Artificial Neural Network (ANN) and SVR algorithms on the time series data of S&P 500 index and SVR outperformed ANN [1]. **Roy et al.** worked on the time series data of Goldman Sachs Group Inc. stocks in 2015 using LASSO based linear regression and the proposed model outperformed the ridge linear regression [2]. **Aseervatham et**

**al.** [3] claimed in 2011 that the ridge logistic regression can achieve the same performance as the Support Vector Machine. **Nair et al.** [4] in 2010, proposed an automated decision tree-adaptive neuro-fuzzy hybrid automated system for stock market prediction. **Gunter, Senyurt. [5]** in 2012 used Random Forest and CART (Classification and Regression tree) for the analysis of ten-year data of technical indicators. **Thakur et al.** developed a stock prices Recommendation system using different regression models i.e: Lasso linear Regression, Ridge Linear Regression, Linear Regression, SVR, KNN and Random Forest and SVR outperformed these models [7]. Similar work has been done on Apples.inc stocks by **Nunno, Lucas** [8]. **Wu et al.** created a recommendation system when to by a stock and when not buy it by introducing a two-layer bias decision tree with technical indicator features [18]. **Huang et al.** proposed a model for the prediction of the stock market by combining the filter rule and decision tree technique [19]. **Yang et al.** studied the asymptotic properties of adaptive elastic net in ultrahigh dimensional sparse linear Regression models and proposed a new model to improve the prediction accuracy [20].

**Gharehchopogh et al.** proposed Linear Regression model for the prediction of the stock market and prove that model has similar performance compared to the real volume by using R square and performance measure [17]. **Sathyaraj, R.** has performed a comparative study of three regression algorithms (Linear Regression, Ridge Regression & SVM) on nine different stocks to predict the next day price. $R^2$ and RMSE have been used to find the correctness of these models [10]. **Kazem, Ahmad, et al.** selected three stocks time series data from NASDAQ to find the correctness of the proposed model. A forecasting model based on SVR, chaotic mapping and firefly algorithm has been proposed for the prediction of stock market prices. MSE and Mean Absolute Percent Error (MAPE) used as performance measures [11]. **Guresen et al.** performed a comparison study for the prediction or daily rate value of NASDAQ stock Index [12]. **Weng et al.** predicted one day ahead movement of NASDAQ Apple stocks using disparate data sources [13]. **Di, Xinjie.** have chosen three datasets i.e Apple Inc., Microsoft Corporation & Amazon.com, inc. and he used extremely randomized tree algorithm for features selection. These features are then fed to SVM for training to predict the next 3-day, next 5-day, next 7-day, next 10-day trend [14].

*B. Details Of Experimental Algorithms*

*1. Support Vector Regression*

Support vector machine is a supervised machine learning algorithm which can be used for classification, regression or outlier detection. It constructs a hyperplane or set of hyperplanes in high dimensional space. SVR model maintains all the main features that characterize the algorithm. It depends only on a subset of training data because of its cost function for the development of the model which ignores any training data close to the model prediction.

*2. Linear Regression*

Linear Regression is predicting the value of a dependent variable Y on some independent variable X provided there is a linear relationship exits. This relationship can be represented by a straight line. For more than one independent variables, the algorithm is called multiple linear Regression.

*3. Lasso Regression*

Least Absolute Shrinkage and Selection Operator (LASSO) is a modification of the Least Square Method which performs very well when the count of features is less as compared to count of observations. It produces solutions by estimating sparse coefficients. It uses L1 norm which is equal to absolute value of the magnitude of coefficients. It performs features selection and shrinkage by reducing coefficients of others to zero.

*4. Ridge Regression*

Ridge Regression is a form of regularized linear regression which performs very well when the count of features is less as compared to the count of observations. It belongs to the class of regression tools which use L2 regularization which adds up L2 penalty which is equals to square of magnitude of coefficients. It can't zero out coefficients thus, it either includes all coefficient or none of them.

*5. Elastic Net Regression*

Elastic Net is a hybrid model with features of both Ridge and LASSO Regression. It works well on a larger dataset. Its uses multicollinearity and regularization values in its optimization function which inherently increase the overfitting and increase the predictive power of the model. It is linearly panelized with both L1 and L2 norms consequence of which it effective shrink some of the coefficients and set some of them to zero.

*6. RANSAC*

RANSAC is a non-deterministic iterative algorithm which estimates parameters from a subset of inliers from the complete dataset that contains outliers. It divides the data into two categories outliers and inliers then use inliers for the model.

*7. Decision Tree*

Decision tree Regressor builds a tree incrementally by splitting the dataset into subsets which results in a tree with decision nodes and leaf nodes. A decision node has two or more branches each representing values for the attribute tested. Leaf node represents the decision on the numerical target. The topmost node is called the root node which corresponds to the best predictor.

*8. Extra Tree*

Extra Tree regressor (stands for extremely randomized Tees) is built differently from the classic decision trees because of its strategy to split nodes. It performs splits for each of the max features randomly and it also selects features randomly and the best split among those is chosen. When max-feature is set to 1, it built a totally decision tree every time.

*9. Random Forest*

Random forest is the association of various binary Regression trees. Independent subset of variables is used for the growth of these large number of binary regression trees. Bootstrapped samples of the dataset build the decision trees and Random forest randomly selects the variables to split [6].

### III. DATASET

The target market for this comparison is NASDAQ because it's not restricted to companies that have U.S. headquarters. Historical data of daily close, open, low and high is gathered from https://www.nasdaq.com. Python library beautifulsoup4 has been used for scrapping data. Our target is prediction of next day opening price.

Target = current day closing price – next day opening price

To show the diversity of this work we randomly choose ten different stocks from five different sectors having 10 years of historical data. Table 1 shows the details of target companies.

| Sector | Industry | Company | symbol |
|---|---|---|---|
| **Finance** | Major Bank | Ameris Bancorp | ABCB |
| | Property casualty insurer | Amerisafe,inc. | AMSF |
| **Technology** | Computer Software | Microsoft Corporation | MSFT |
| | Computer Manufacturing | Apple Inc. | AAPL |
| **Energy** | Coal Mining | Alliance Holdings GP, LP. | AHGP |
| | Industrial Machinery/ Component | American Electric Technologies, NC. | AETI |
| **Health Care** | Major Pharmaceuticals | ACADIA Pharmaceuticals Inc. | ACAD |
| | Medical/ Dental instruments | Accuracy incorporated | ARAY |
| **Consumer services** | Other Specialty Stores | 1-800Flowers.com, inc | FLWS |
| | Catalog/ Specialty Distribution | Amazon.com, inc. | AMZN |

Table 1: Name of stocks along with symbols

The data size of each stock is 2519 sample points. The size of the training set is 80% of the stock and rest is used for testing.

### IV. EXPERIMENTAL SETUP

Python 3.5.2 along with beautifulsoup4, numpy 1.12.1, pandas 0.19.2, urllib, requests, parse, re used on Ubuntu 16.04 for experimental setup. For training and finding the polarity of a test data set nine different supervised machine learning regressors has been used. Scikit-learn 0.17 is used for training followed by prediction on the dataset. This tool provides reusability in multiple contexts and is easy to use with multiple machines learning regressors

### V. PERFORMANCE MEASURE

There are many evaluation metrics that exist for evaluating prediction on regression Machine Learning algorithms. To evaluate our models, Mean Square Error(MSE) and $R^2$ Error have been used.

*C. Mean Square Error*

MSE provides the idea of magnitude of an error. It calculates the average of the square of the errors which is the difference between the target value and the predicted value. It is also considered as a risk function which measures the quality of an algorithm. It is always non-negative closer the value to zero better the Algorithm.

$$MSE(Target, Predicted) = \frac{1}{N} \sum (target-predicted)^2$$

*Where N is the total numbers of sample points.*

*D. $R^2$ Error*

R square is also called coefficient of determination. It provides the measures of how well the algorithm likely to be predicted for the future. In the worst case, the model likely to perform negatively. The value between 0 and 1 represents no fit and best fit respectively. Higher the value better the Algorithm.

$$R^2(Target, Predicted) = 1 - \frac{\sum (target-predicted)^2}{\sum (target-target\ Mean)^2}$$

### VI. RESULTS AND DISCUSSIONS

we have compared 9 different algorithms on the data of 10 different companies each belongs to a different industrial sector to find out which regression algorithm suits best for these kinds of datasets so that researcher find it helpful for their future work. Even for a beginner it will be helpful and easy to understand. Table 2 represents results of these algorithms for all the stocks. Each column represents a stock and each row represents MSE and $R^2$ value.

When these algorithms applied on the data of Microsoft Corporation (MSFT) a technology company, Decision Tree and Extra Tree showed best results on training data with 0.0000 MSE and 1.000 $R^2$ while Lasso and Ransac performed worst with MSE 0.1175 & 0.1344 and $R^2$ 0.000 & -0.1444 respectively. On the other hand, SVM and Lasso showed best results on test data with MSE 0.2642 & 0.2639 and $R^2$ -0.0078 & -0.0066 while Ransac and linear regression performed worst on test data with MSE 0.4951 & 0.4672 and $R^2$ -0.0291 & -0.8887 respectively.

On the data of Apple inc. (AAPL) a technology company, Decision Tree and Extra Tree Algorithms showed best results

on training data with 0.0000 MSE and 1.0000 $R^2$ while Lasso and Ransac performed worst with MSE 0.6624 & 0.6769 and $R^2$ 0.0002 & -0.0217 respectively. Similarly, ElasticNet and SVM showed best results on test data with MSE 1.1796 & 1.1801 and 0.1582 & -0.0017 $R^2$, on the other hand linear regression and Decision Tree performed worst on test data with MSE 2.5634 & 1.6138 and -0.0294 & -0.3698 $R^2$.

While training the algorithms with the data of Amerisafe inc. (AMSF) a finance company, Decision Tree and Extra Tree showed best results with 0.0000 MSE and 1.0000 $R^2$ while Linear Regression and Ransac performed worst with 0.4339 & 0.1133 MSE and 0.0057 & -0.1528 $R^2$ respectively. While Testing SVM and Lasso showed best results with MSE 0.1708 and -0.0000 $R^2$, Ransac `and linear regression performed worst with MSE 0.2032 & 0.4339 and -0.1901 & -0.0007 $R^2$.

Among all the algorithms applied on the data of Ameris Bancorp (ABCB) a finance company, Decision Tree and Extra Tree performed best on training data with 0.0000 MSE and 1.0000 $R^2$ while Linear Regression and Ransac performed worst with 0.0499 & 0.0368 MSE and 0.0059 & -0.4673 $R^2$ respectively. Linear Regression and Lasso showed best results on test data with MSE 0.0996 & 0.1923 and -0.0181 & -0.0056 $R^2$, Decision Tree `and Extra Tree performed worst on test data with mse 0.2146 & 0.1923 and -0.5436 & -0.3826 $R^2$.

On the application of these algorithms on the data of Alliance Holdings GP, LP (AHGP) a coal mining company, Decision Tree and Extra Tree showed best results on training data with 0.0000 MSE and 1.0000 $R^2$ while Linear Regression and Ransac performed worst with 0.00481 & 0.0265 MSE and 0.0059 & -0.0517 $R^2$ respectively. while linear Regression and Lasso showed best results on test data with MSE 0.0996 & 0.1398 and -0.0181 & -0.0056 $R^2$ and Decision Tree and Extra Tree performed worst on test data with MSE 0.2300 & 0.1946 and -0.6543 & -0.3994 $R^2$.

Decision Tree and Extra Tree showed best results on training data of American Electric Technologies, inc. (AEIT) a industrial machinery company, with 0.0005 MSE and 0.9681 $R^2$ while Linear Regression and Ransac performed worst with 0.00487 & 0.0179 MSE and 0.0470 & -0.2006 $R^2$ respectively. On the other hand, Ridge Regression and Lasso showed best results on test data with MSE 0.0057 & 0.0059 and -0.0822 & -0.1357 $R^2$ and Decision Tree `and Linear Regression performed worst on test data with MSE 0.0433 & 0.0567 and -7.2923 & -0.0804 $R^2$.

When these algorithms trained with the data of ACADIA Pharmaceuticals Inc. (ACAD) a major pharmaceutical company, Decision Tree and Extra Tree showed best results with 0.0000 MSE and 1.0000 $R^2$ while Lasso Regression and Ransac performed worst with 0.1478 & 0.1594 MSE and 0.0000 & -0.0786 $R^2$ respectively. Similarly, ElasticNet Regression and Lasso showed best results on test data with MSE 0.2904 & 0.2910 and -0.0013 & -0.0034 $R^2$ and Decision Tree `and Linear Regression performed worst on test data with MSE 0.5316 & 0.5728 and -0.8331 & -0.0170 $R^2$.

To perform the comparison of these algorithms for the data of Accuracy incorporated (ARAY) a medical and dental instrument company, Decision Tree and Extra Tree showed best results on training data with 0.0000 MSE and 1.0000 $R^2$ while Linear Regression and Ransac performed worst with 0.0897 & 0.0126 MSE and 0.0038 & -0.1031 $R^2$ respectively. On the other hand, ElasticNet Regression, Lasso & Ridge showed best results on test data with MSE 0.0059 and $R^2$ -0.0019, -0.0016 & 0.0002 and Decision Tree `and Linear Regression performed worst on test data with MSE 0.0097 & 0.0455 and -0.5301 & 0.0007 $R^2$.

On training data of 1-800Flowers.com, inc (FLWS) other specialty store, among all algorithms Decision Tree and Extra Tree showed best results with 0.0000 MSE and 1.0000 $R^2$ while Ransac with 0.0062 MSE and -0.0729 $R^2$ SVM, Linear Regression, Ridge & Lasso performed worst with 0.0057 MSE and 0.0039, 0.0036, 0.0036 & 0.0000 $R^2$ respectively. On the other hand, ElasticNet Regression, Lasso, SVM, Linear Regression & Ridge showed best results on test data with MSE 0.0093 and -0.0014, -0.0000, -0.0096 & -0.0070 $R^2$ and Decision Tree and Extra Tree performed worst on test data with MSE 0.0242 & 0.0131 and -1.6146 & -0.4154 $R^2$.

While applying these algorithms on the data of Amazon.com, inc. (AMZN) a Catalog/ Specialty Distribution, Decision Tree and Extra Tree showed best results on training data with 0.0000 MSE and 1.0000 $R^2$ while ElasticNet Regression and Ransac performed worst with 20.2486 & 22.2728 MSE and 0.0066 & -0.0927 $R^2$ respectively. On the otherhand, ElasticNet Regression and Lasso showed best results on test data with MSE 83.1652 & 83.2893 and 0.0070 & 0.0056 $R^2$ and Decision Tree `and Random Forest performed worst on test data with MSE 106.4541 & 206.5049 and $R^2$ -0.2710 & -1.4656.

Over all Decision Tree and Extra Tree are best fit on training data for all the companies where as Linear regression, Lasso and RANSAC are worst fit. SVR, Lasso, ElasticNet and ridge are best fit on testing data for all the companies where as Linear regression, Ransac, Decision Tree and Extra Tree are worst fit. Due to the overfitting nature of the Decision tree and Extra Tree they have this tendency to perform good on training data but unable to perform well on testing data. Ransac doesn't performed well on both training and testing data due to its unsuitability for time series data. Similarly, Linear regression implemented on the base of statistical model that, shows optimal results when the relationships between the independent variables and the dependent variable are almost linear. SVR is not biased by outliers and easily adaptable, it works very well on non-linear problems. Lasso shrink's coefficients exactly to zero, which performs the feature selection using L1 regularization. Ridge diminishes the value of coefficients but does not reach zero, which shows the no feature selection feature. Lasso and Ridge allows to use complex models and avoid over-fitting simultaneously. ElasticNet allows to extend some of Ridge's stability under rotation and there is no limit to the number of selected variables.

| Model name | PM | MSFT | AAPL | AMSF | ABCB | AHGP | AETI | ACAD | ARAY | FLWS | AMZN |
|---|---|---|---|---|---|---|---|---|---|---|---|
| Support Vector Regression | mseTrain | 0.1156 | 0.0085 | 0.0316 | 0.0052 | 0.0052 | 0.0143 | 0.1309 | 0.0114 | 0.0057 | 17.5923 |
| | mseTest | 0.2642 | 1.1801 | 0.1708 | 0.1402 | 0.1402 | 0.0077 | 0.2986 | 0.0060 | 0.0093 | 86.6440 |
| | $R^2$ Train | 0.0158 | 0.9872 | 0.6782 | 0.7915 | 0.7915 | 0.0366 | 0.1148 | 0.0031 | 0.0039 | 0.1369 |
| | $R^2$ Test | -0.0078 | -0.0017 | -0.0000 | -0.0084 | -0.0084 | -0.4698 | -0.0297 | -0.0042 | -0.0096 | -0.0345 |
| Linear Regression | mseTrain | 0.0813 | 0.0813 | 0.4339 | 0.0499 | 0.0481 | 0.0487 | 0.0042 | 0.0897 | 0.0057 | 20.1704 |
| | mseTest | 0.4672 | 2.5634 | 0.3906 | 0.0996 | 0.0996 | 0.0567 | 0.5728 | 0.0455 | 0.0093 | 83.3354 |
| | $R^2$Train | 0.0047 | 0.0323 | 0.0057 | 0.0059 | 0.0059 | 0.0470 | 0.0307 | 0.0038 | 0.0036 | 0.0105 |
| | $R^2$Test | -0.0291 | -0.0294 | -0.0007 | -0.0181 | -0.0181 | -0.0804 | -0.0170 | 0.0007 | -0.0070 | -0.0050 |
| Ridge Regression | mseTrain | 0.1169 | 0.6412 | 0.0977 | 0.0249 | 0.0249 | 0.0142 | 0.1433 | 0.0114 | 0.0057 | 20.1704 |
| | mseTest | 0.2697 | 1.2125 | 0.1709 | 0.1415 | 0.1415 | 0.0057 | 0.2948 | 0.0059 | 0.0093 | 83.3354 |
| | $R^2$Train | 0.0047 | 0.0323 | 0.0057 | 0.0059 | 0.0059 | 0.0469 | 0.0307 | 0.0038 | 0.0036 | 0.0105 |
| | $R^2$Test | -0.0289 | -0.0292 | -0.0007 | -0.0180 | -0.0180 | -0.0822 | -0.0166 | 0.0002 | -0.0070 | 0.0050 |
| Lasso Regression | mseTrain | 0.1175 | 0.6624 | 0.0983 | 0.0251 | 0.0251 | 0.0149 | 0.1478 | 0.0114 | 0.0057 | 20.2831 |
| | mseTest | 0.2639 | 1.1805 | 0.1708 | 0.1398 | 0.1398 | 0.0059 | 0.2910 | 0.0059 | 0.0093 | 83.2893 |
| | $R^2$Train | 0.0000 | 0.0002 | 0.0000 | 0.0000 | 0.0000 | 0.0000 | 0.0000 | 0.0000 | 0.0000 | 0.0049 |
| | $R^2$Test | -0.0066 | -0.0020 | -0.0000 | -0.0056 | -0.0056 | -0.1357 | -0.0034 | -0.0016 | -0.0000 | 0.0056 |
| ElasticNet Regression | mseTrain | 0.1174 | 0.6545 | 0.0981 | 0.0250 | 0.0250 | 0.0149 | 0.1462 | 0.0114 | 0.0057 | 20.2486 |
| | mseTest | 0.2675 | 1.1796 | 0.1711 | 0.1410 | 0.1410 | 0.0062 | 0.2904 | 0.0059 | 0.0093 | 83.1652 |
| | $R^2$Train | 0.0004 | 0.0121 | 0.0018 | 0.0008 | 0.0008 | 0.0020 | 0.0108 | 0.0002 | 0.0028 | 0.0066 |
| | $R^2$Test | -0.0203 | 0.1582 | -0.0020 | -0.0142 | -0.0142 | -0.1830 | -0.0013 | -0.0019 | -0.0014 | 0.0070 |
| Random Forest Regression | mseTrain | 0.0233 | 0.1582 | 0.0263 | 0.0055 | 0.0055 | 0.0037 | 0.0260 | 0.0027 | 0.0012 | 3.6827 |
| | mseTest | 0.3238 | 1.4956 | 0.1745 | 0.1846 | 0.1846 | 0.0098 | 0.3368 | 0.0066 | 0.0115 | 206.5049 |
| | $R^2$Train | 0.8019 | 0.7612 | 0.7319 | 0.7809 | 0.7809 | 0.7482 | 0.8244 | 0.7637 | 0.7994 | 0.8193 |
| | $R^2$Test | -0.2351 | -0.2695 | -0.0217 | -0.3279 | -0.3279 | -0.8821 | -0.1612 | -0.1108 | -0.2429 | -1.4656 |
| RANSAC | mseTrain | 0.1344 | 0.6769 | 0.1133 | 0.0368 | 0.0265 | 0.0179 | 0.1594 | 0.0126 | 0.0062 | 22.2728 |
| | mseTest | 0.4951 | 1.3181 | 0.2032 | 0.1653 | 0.1560 | 0.0060 | 0.3093 | 0.0064 | 0.0107 | 102.5914 |
| | $R^2$Train | -0.1444 | -0.0217 | -0.1528 | -0.4673 | -0.0571 | -0.2006 | -0.0786 | -0.1031 | -0.0729 | -0.0927 |
| | $R^2$Test | -0.8887 | -0.1188 | -0.1901 | -0.1890 | -0.1221 | -0.1464 | -0.0666 | -0.0806 | -0.1545 | -0.2249 |
| Decision Tree Regression | mseTrain | 0.0000 | 0.0000 | 0.0000 | 0.0000 | 0.0000 | 0.0005 | 0.0000 | 0.0000 | 0.0000 | 0.0000 |
| | mseTest | 0.3236 | 1.6138 | 0.1851 | 0.2146 | 0.2300 | 0.0433 | 0.5316 | 0.0091 | 0.0242 | 106.4541 |
| | $R^2$Train | 1.0000 | 1.0000 | 1.0000 | 1.0000 | 1.0000 | 0.9681 | 1.0000 | 1.0000 | 1.0000 | 1.0000 |
| | $R^2$Test | -0.2344 | -0.3698 | -0.0838 | -0.5436 | -0.6543 | -7.2923 | -0.8331 | -0.5301 | -1.6146 | -0.2710 |
| ExtraTree Regression | mseTrain | 0.0000 | 0.0000 | 0.0000 | 0.0000 | 0.0000 | 0.0005 | 0.0000 | 0.0000 | 0.0000 | 0.0000 |
| | mseTest | 0.3250 | 1.4562 | 0.1994 | 0.1923 | 0.1946 | 0.0169 | 0.3585 | 0.0069 | 0.0131 | 104.0198 |
| | $R^2$Train | 1.0000 | 1.0000 | 1.0000 | 1.0000 | 1.0000 | 0.9681 | 1.0000 | 1.0000 | 1.0000 | 1.0000 |
| | $R^2$Test | -0.2399 | -0.2361 | -0.1676 | 0.3826 | -0.3994 | -2.2333 | -0.2363 | -0.1688 | -0.4154 | -0.2420 |

Table 2: Results of Algorithms for All stocks

### VII. CONCLUSION

This paper presents a comparitive study of different machine learning Regression models for the prediction of next day opening prices. The overwhelming majority of research in this field, using all techniques, has focused on forecasting the price path in the future. This suggests that the majority of researchers were mostly interested in determining if the market price will rise or fall in the future. The margin of rise or decrease in price is not revealed by determining the path. The findings indicate that all three strategies are capable of forecasting the index's price. Forecasting the price is much more difficult than forecasting the route. It will provide more detail regarding the potential behavior of the market stock price if it is forecast with a certain degree of precision. This sends a strong signal to investors that demand forecasts may be done using computational intelligence techniques. Another aim of the research was to show that stock values in the past can be used to forecast future prices. Past stock markets provide details that can be used to forecast future prices, according to the findings. A portfolio of ten different stocks from five different sectors of industries has been built to find out the best suitability of these models. To Evaluate these models MSE and $R^2$ is used as performance measures. From above discussion, it is concluded that SVR, Lasso, ElasticNet and ridge regression has higher accuracy as compared to Decision Tree, Extra Tree and Ransac.